\documentclass{vienna-conf2019}
\usepackage{graphicx}
\usepackage{hyperref}
\usepackage[]{natbib}  
\usepackage{epstopdf}

\def\BibTeX{{\rm B\kern-.05em{\sc i\kern-.025em b}\kern-.08em
    T\kern-.1667em\lower.7ex\hbox{E}\kern-.125emX}}
\bibpunct{(}{)}{;}{a}{}{,}  

\begin{document}

\TitreGlobal{Stars and their variability observed from space}


\title{The BRITE-SONG of Aldebaran - Stellar Music in three voices}

\runningtitle{The three voices of the BRITE-SONG of Aldebaran}

\author{P.\,G.\,Beck$^{1,2,}$}
\address{Institute of Physics, University of Graz, NAWI Graz, Universitätsplatz 5/II, 8010 Graz, Austria}
\address{Instituto de Astrofísica de Canarias, 38205 La Laguna, Tenerife, Spain}
\address{Departamento de Astrofísica, Universidad de La Laguna, 38206 La Laguna, Tenerife, Spain}
\author{R.\,Kuschnig}\address{Inst.\,of\,Communication\,Networks\,\&\,Satellite\,Communications, Graz\,University\,of\,Technology, Infeldgasse\,12, 8010\,Graz, Austria}
\author{G.\,Houdek}
\address{Aarhus University, Department of Physics and Astronomy, Ny Munkegade 120, 8000 Aarhus C, Denmark}

\author{T.\,Kallinger}\address{Institut für Astrophysik der Universität Wien, Türkenschanzstr. 17, 1180 Vienna, Austria}
\author{W.\,W.\,Weiss$^6$}
\author{P.\,L.\,Palle$^{2,3}$}
\author{F.\,Grundahl$^5$}
\author{A.\,Hatzes}\address{Thüringen Landessternwarte Tautenburg, Sternwarte 5, D-07778 Tautenburg, Germany;}
\author{H.\,Parviainen$^{2,3}$}
\author{C.\,Allende Prieto$^{2,3}$}
\author{H.\,J.\,Deeg$^{2,3}$}
\author{A.\,Jim{\'e}nez$^{2,3}$}
\author{S.\,Mathur$^{2,3}$}
\author{R.\,A.\,Garcia}\address{AIM, CEA, CNRS, Universite Paris-Saclay, Universite Paris Diderot, Sorbonne Paris Cite, F-91191 Gif-sur-Yvette, France}
\author{T.\,R.\,White$^5,$}\address{Sydney Institute for Astronomy (SIfA), School of Physics, University of Sydney, NSW 2006, Australia}
\author{T.\,R.\,Bedding$^{9,5}$}
\author{D.\,H.\,Grossmann$^{1}$} 
\author{S.\,Janisch$^{1}$} 
\author{T.\,Zaqarashvili$^{1}$}
\author{A.\,Hanslmeier$^{1}$} 
\author{K.\,Zwintz}\address{ Institut fur Astro- und Teilchenphysik, Universität Innsbruck, Technikerstrasse 25, A-6020 Innsbruck, Austria}
\author{the BRITE  \& SONG teams}





\setcounter{page}{1}


\maketitle
\begin{abstract}
Solar-like oscillations in red-giant stars are now commonly detected in thousands of stars with space telescopes such as the NASA \textit{Kepler} mission. Parallel radial velocity and photometric measurements would help to better understand the physics governing the amplitudes of solar-like oscillators. Yet, most target stars for space photometry are too faint for light-demanding ground-based spectroscopy. The BRITE Constellation satellites  provide a unique opportunity of two-color monitoring the flux variations of bright luminous red giants. Those targets are also bright enough to be monitored with high-resolution spectrographs on small telescopes, such as the SONG Network. In these proceedings we provide a first overview of our comprehensive, multi-year campaign utilizing both BRITE and SONG to seismically characterize Aldebaran, one of the brightest red giants on the sky.
Because luminous red giants can be seen at large distances, such well characterized objects will serve as benchmark stars for galactic archeology.
\end{abstract}



\section{Introduction \label{beck:sec:intro}}
Space telescopes such as the NASA \textit{Kepler} or TESS missions \citep[respectively]{Borucki2010, Ricker2015} have allowed for the detection of solar-like oscillations in ten-thousands of main-sequence and red-giant stars \citep[e.g.][]{Hon2018, Garcia2019,SilvaAguirre2019}. While such satellites provide ultra-precise monochromatic photometry, typical target stars are too faint for light-demanding ground-based complementary techniques. Simultaneous multi-color photometry and parallel radial-velocity (RV) monitoring of solar-like oscillating stars, however, would provide crucial information for understanding the physics governing the oscillations and their amplitudes. So far, the only solar-like oscillators for which such simultaneous data have been acquired and analysed are the Sun itself \citep{no5} and the sub-giant \hbox{Procyon \citep{Arentoft2008}.}

The 3\,cm telescopes of the five BRITE Constellation satellites \citep[BRIght Target Explorer,][]{no10}, with its blue and red photometric filters, have enabled multi-color space photometry of very bright targets since the launch of the the first pair of satellites in 2013. While the primary science case for BRITE satellites does not include solar-like oscillations on the main sequence, or evolved stars, it was shown by \cite{Kallinger2019} that red-giant stars with oscillation frequencies below 10\,$\mu$Hz (R$_\star$$>$$\sim$25R$_\odot$) exhibit oscillation amplitudes that are large enough to be detected by the BRITE satellites. Such targets are also accessible by spectrographs mounted on 1m-class telescopes, such as SONG \citep[Stellar Observations Network Group,][]{Grundahl2017}, which is designed for the acquisition of high-quality spectroscopic time series for asteroseismology. The first telescope of this emerging telescope network was commissioned 2014, which is installed at the Teide observatory on Tenerife, and is equipped with a high-resolution spectrograph (with a spectral resolution of 77000\,$\leq$ R $\leq$ 112000) and the capability of simultaneous wavelength calibration providing meter-per-second RV precision.

In this project we are using the combination of multi-color BRITE photometry and SONG spectroscopy to investigate solar-like oscillations in the luminous red-giant branch star Aldebaran ($\alpha$\,Tau) with a visual brightness of +0.9\,magnitudes and a stellar luminosity of 439$\pm$17\,L$_\odot$ \citep{Heiter2015}. Photometry obtained by the \textit{Kepler} K2 mission by \cite{no6} for this star shows intensity variations of $\sim$5.7days, which correspond to an oscillation frequency of $\sim$2 \,$\mu$Hz. Their analysis concluded that this star has a mass of 1.16$\pm$0.07\,M$_\odot$.
 Furthermore, interferometric radii are available \citep{Richichi2005} providing the opportunity to test seismic scaling relations for luminous RGBs, which are suspected to significantly depart from the well-established seismic scaling relations \citep{Tu2020}.
  More technical details are provided in Secion\,\ref{beck:sec:obs}. From more than three decades of RV measurements \citep{no7} have revealed the presence a quasi-periodic modulation with a period of $\sim$700 days, which is not reflected in the variation of the strength of the emission lines of the core of Ca\,H\&K lines, which are a classical activity indicator in cool stars. The authors suggested that this signal could originate from a planetary companion of 6 Jupiter masses but also urged caution as the signal is not stable over decades as expected for a planet, and over-stable convection could serve as an alternative explanation. 
As we will discuss further in Section\,\ref{beck:sec:planet}, this planet candidate has been heavily debated in the literature.

\section{Variations in the intensity vs. velocity field \label{beck:sec:theo}}

\begin{figure}[t!]
 \centering
\includegraphics[width=0.8\linewidth]{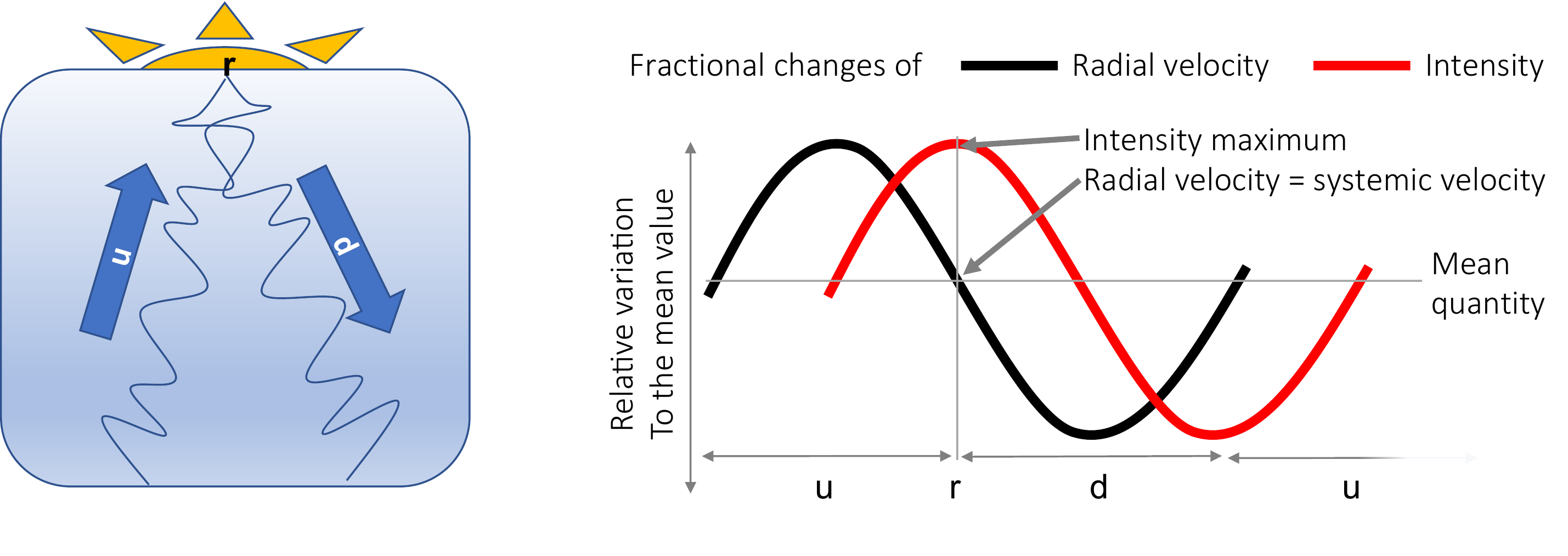}      
\vspace{-7mm}  \caption{Conceptual illustration of the behaviour of an oscillation mode near the surface of a star. The left panel depicts a wave coming from the interior (u) is reaching regions of lower density and get reflected (r) back into the deeper and denser layers again (d). In the right panel, two sine curves approximate the variations radial velocity and intensity field (black and red, respectively). This sketch assumes ideal adiabatic conditions in the outer atmosphere, with the velocity field leading to the intensity variation shifted by 90 degrees.}\vspace{-3mm} 
  \label{beck:fig:HRD}
\end{figure}

When an oscillation mode reaches the near surface layers, it is reflected back to deeper, denser layers. At the surface, such modes lead to periodic distortions of the surface temperature and velocity field, typically described by the degree $\ell$ of the spherical harmonics 
\cite[see][and references therein]{Aerts2010}. The surface temperature variations lead to fractional variations of the stellar luminosity. Spectroscopy measures the surface field component along the line of sight. Therefore, oscillation modes observationally manifest themselves as variations of the mean brightness and the systemic radial velocity of the star, respectively. At the reflection point an oscillation mode exhibits its maximum brightness, at it is least obscured by overlying layers. Contemporaneously, the perturbation of the velocity field is changing its propagation direction. Therefore, the velocity field for a
given mode shows no perturbation and equals the systemic radial velocity of the star. As illustrated in Figure \ref{beck:fig:HRD}, this leads to a phase shift of -90\,deg in the ideal case.

The two parameters describing the differences between the velocity and intensity variations in Fourier space are the $(i)$ amplitude ratio and $(ii)$ the phase difference. The quantities provide two important diagnostic constraints.
The information contained in these parameters allows us to test stellar oscillation and atmospheric models beyond the possibilities of classical asteroseismology. Nevertheless, due to the lack of simultaneous appropriate data, such studies have only been carried out for two solar-like oscillators.

As demonstrated by \cite{no1} and \cite{no3}, a calibrated amplitude ratio between intensity and velocity variations provides strong excitation-model independent constraints on the stellar atmosphere. While the RV amplitude is determined by the velocity fluctuations from oscillations and granulation along the line of sight, and therefore allows a direct comparison with models,  the case of photometric variations remains challenging \cite{no2}. The intensity variations correspond primarily to the temperature fluctuations of the atmosphere over the oscillation cycle, but need to be translated into fractional changes  of the bolometric luminosity. However, for a comparison  between observations and models, observational aspects such as the photometric passband and the color dependent quantum efficiency of the detector of the space telescope need to be taken into account. In this respect, simultaneous multi-color photometry provides different measures to the same quantity and increases the robustness of the experiment.

The phase shift between oscillation modes seen in these two observables is a relevant parameter for constraining pulsation eigenfunctions and eventually models for stochastic excitation \citep{no3}. Were the oscillations purely adiabatic, the RVs are expected to lead the intensity variations with a phase shift of -90\,deg, when normalised by the oscillation period. The solar case has been found to be close to this value \citep{no5}. Any departure from this value reveals heat transfer between the stellar background and the oscillation modes \citep{no4,no5}, with a maximal lag of -180\,deg in the isothermal case.

\section{Experimental data: 2016-2020 \label{beck:sec:obs}}
\begin{figure}[t!]
 \centering
\includegraphics[width=1\linewidth, height=72mm
]{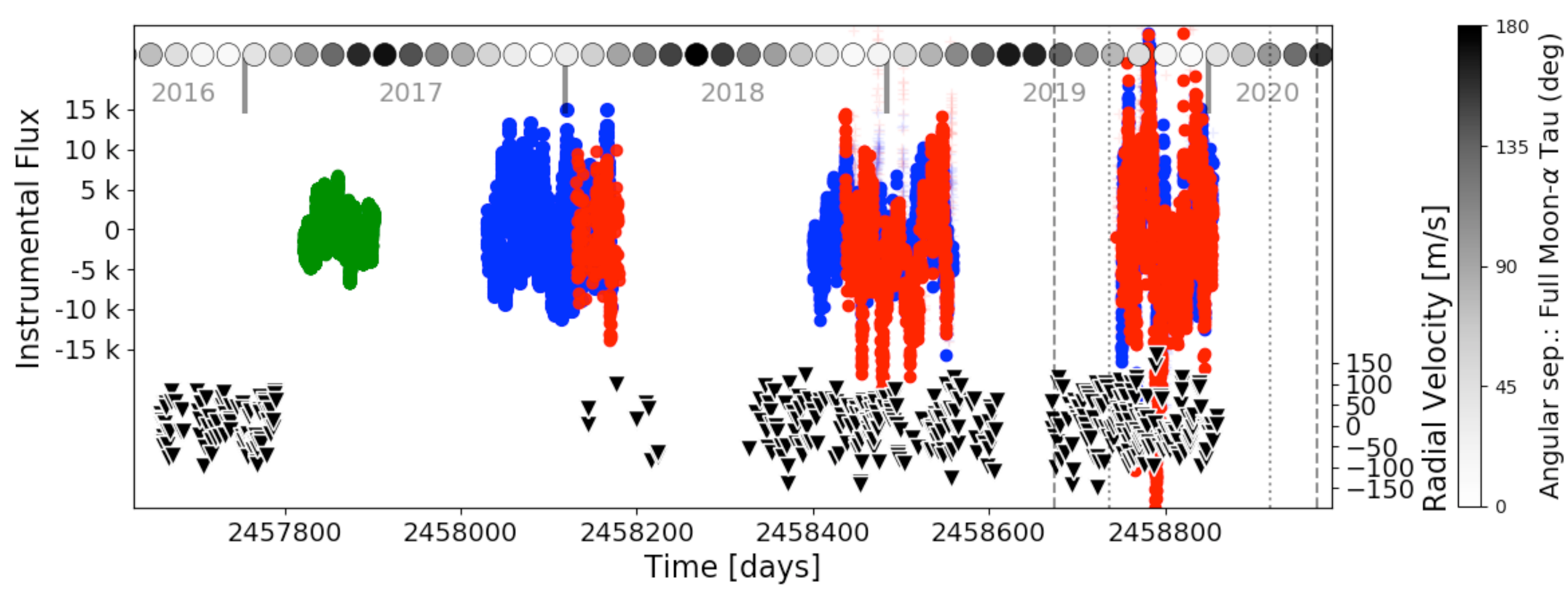}      
\vspace{-2mm}  
\caption{Data of the ongoing observing campaign on Aldebaran. 
  The red and blue points depict the orbital mean values of the measured photometric flux by the blue filtered BRITE/Lem and the red filtered BRITE/Toronto satellites, respectively. The 80 days of K2 photometry \citep{no6} are shown in green. The radial velocities, measured with SONG are depicted as black triangles, whereby the long-periodic trend has been removed. The duration of the ongoing SONG and BRITE observations in the season 2019/20 are indicated by dashed and dotted lines respectively. On top of the data the years as well as 
 the angular distance between the star and the moon at the moment of the plenilunium is indicated with the scale of grays given to the right. Due to stray-light contamination, data points taken with the moon closer than 35 degrees are shown as semi-transparent crosses. }
  \label{beck:fig:campaign}
\end{figure}

The multi-technique campaign presented in this work (see Figure\,\ref{beck:fig:campaign}) was started in visibility season 2017/18 of Aldebaran with the observations with BRITE/Lem, which is equipped with a blue photometric filter (400-450\,nm). At the end of the same observing season, this star was also added to the observing program of BRITE/Toronto, which is equipped with a red filter (500-700\,nm). Each data point shown in Figure\,\ref{beck:fig:campaign} represents the mean of all the individual measurements obtained during a satellite orbit, whereby we discard orbits with less than 15 individual measurements. The orbital period of the BRITE satellites is about 90 minutes, resulting in 14 to 15 photometric measurements per day.

Towards the end of this season, we also attempted simultaneous spectroscopic observations with the SONG telescope. The standard observing program forsees for 1 or 2 pointings a night with three consecutive spectra with a resolution of 90.000 and an Iodine-cell for simultaneous wavelength calibration. However, we were struck with adverse weather conditions that were extreme  in intensity and duration for Tenerife. 

Since observing season 2018/19, all three instruments are well coordinated. A typical observing season with BRITE lasts for about 140 days (September - March), while SONG can obtain for about 300 days in a row at least one set of spectra per night. This time series is only interrupted between May and July. Periodic gaps are introduced in the dataset by the moon. Given the low ecliptical latitude of Aldebaran of about 5\,deg, the photometric observations are contaminated and eventually interrupted by the full moon once every $\sim$28 days. We note that stray light from the moon is not an issue for the spectroscopic observations.

The full dataset of the ongoing campaign is depicted in Figure\,\ref{beck:fig:campaign}. This figure also shows the complementary dataset of 80 days with K2 photometry, obtained by \cite{no6} using the photometric technique of Halo-photometry to counteract the effects of partial saturation \cite{White2017}. In 2016 and prior to K2 photometry, \cite{no6} obtained 125\,days of RV with SONG, which we also include in our~analysis.



\section{The controversy on the planet orbiting Aldebaran \label{beck:sec:planet}}
The possibility of the presence of a planetary companion around Aldebaran was originally proposed by \cite{no7}. These authors interpreted the long-periodic trend present in the radial velocity signal as a possible indication of a massive Jupiter-like planetary companion. However, they were cautious of firmly declaring it a planet, as the radial velocity signal was not as stable as it would be expected for an exoplanet. They also considered as a possible solution that this signal originated from overstable convection, which could occur under extreme non-adiabatic conditions in the stellar atmosphere \citep{Saio2015}. 

Three years later, parallel to the start of our campaign, numerous studies dedicated to this star or the nature of the long-periodic variations were published. First, \cite{Hatzes2018} showed that the radial velocities in the luminous red-giant branch star $\gamma$ Draconis initially exhibited a similar behaviour. 
However, the RV signals disappeared in the years 2013-16, 
and then reappeared with a phase-shift,
being incompatible with 
a planetary origin.
The authors concluded that this finding also points towards a non-planetary explanation for the long-periodic signal in Aldebaran.
\cite{no6}, who presented a first seismic analsysis based on K2 photometry and SONG RVs, used their results to improve the parameters of the planetary companion. They further argued that Aldebaran\,b as well as $\gamma$ Draconis\,b must be planets, with the argument 
that \textit{"it would be a cruel conspiracy of nature if red giants support a type of oscillation that is common and closely resembles a planetary signal. We believe this cannot be the case"}, \citep[incipit of paragraph 3, Appendix D]{no6}.
Finally, \cite{no8} presented a stability analysis of the dataset of \cite{no7}, complemented with their own spectroscopic observations from Lick Observatory. They tested a two-planet model, which significalty reduced the large RV scatter in the residuals, but found that such solution is very unlikely to be dynamically stable.

\section{Current status of the analysis, conclusions \& Outlook\label{beck:sec:conclusions}}

The ongoing photometric and spectroscopic observations obtained by the BRITE satellites and the SONG telescope are building an unique, multi-year data set to study amplitude and phase differences between velocity and intensity variations. Combining the RV from SONG with multi-color space photometry from the BRITE satellite offers a rare possibility of characterizing the outer most layers of luminous red- giant branch stars and provide crucial diagnostics to test the physics of the stellar atmosphere. These parameters give access to layers in the star that are not well probed by normal oscillation modes. 
Luminous red-giant branch stars like Aldebaran are of further interest as they allow us to study the asteroseismic scaling relations and the stellar structure of objects with increasing departure from the adiabatic conditions in the stellar atmosphere \citep{Tu2020,Kallinger2018}.

As visible in Figure\,\ref{beck:fig:campaign}, the observing project is still ongoing and we only show preliminary data for a sneak analysis. This has in particular an effect on the photometric amplitudes. Therefore, we refrain from quantifying the discussed parameters in these proceedings.

The comparison of photometric and spectroscopic datasets shows the variation in velocity is clearly leading the intensity variation.  Visual inspection in the time domain suggests a phase difference much larger than the -90 degrees of the adiabatic case. This is not surprising, but one needs to be cautious as the interference of granulation, which has comparably large amplitudes and periods as the oscillation signal in Aldebaran, could lead to an overestimated phase difference. Therefore, the final value will be determined from the frequency domain. However, the frequency resolution of a full season of BRITE observations is not sufficient to resolve individual oscillation modes.

The long-periodic modulations previously reported in the literature are also present in the SONG dataset. Yet, it is too early to arrive at concrete conclusions on their origin. The long timebase and the high sampling rate of the radial-velocity measurements will help to decide if these variations are caused by a planetary companion, or by an unidentified physical process. We note that these variations are on different times scales than were found for secondary clump stars of the Hyades, which are likely to originate from rotational modulation \citep{Beck2015,Arentoft2019}. 

Constraining the origin of the signal is essential to be able to disentangle actual planets around red-giant stars from spurious planet detections due to intrinsic effects and activity. The reported observations clearly demonstrate that a firm detection of planets around red-giant stars requires radial-velocity monitoring extending over at least several decades


Understanding such benchmark stars like  Aldebaran, $\gamma$ Draconis or Arcturus  is  of timely relevance, since luminous red giants can even be seen at the deep end of the Milky Way \citep[e.g.][]{Mathur2016}.  Therefore, such luminous red giant stars serve as highly needed probes for understanding extreme phases of stellar evolution, the distribution of exoplanets and eventually the evolution history of our Galaxy.\vspace{-2mm}

\begin{acknowledgements}
The authors gratefully acknowledge the work of the BRlTE-constellation and SONG science and technical teams. 
Based on data collected by the BRITE Constellation satellite mission, designed, built, launched, operated and supported by the Austrian Research Promotion Agency (FFG), the University of Vienna, the Technical University of Graz, the University of Innsbruck, the Canadian Space Agency (CSA), the University of Toronto Institute for Aerospace Studies (UTIAS), the Foundation for Polish Science \& Technology (FNiTP MNiSW), and National Science Centre (NCN).
We would like to acknowledge the Villum Foundation, The
Danish Council for Independent Research | Natural Science,
and the Carlsberg Foundation for the support on building the
SONG prototype on Tenerife.
We thank the SOC/LOC of the \textit{"Stars and their Variability"} conference for organizing an inspiring meeting. 
\end{acknowledgements}

\bibliographystyle{aa}  
\bibliography{beck_3o01.bib} 

\end{document}